\documentclass[12pt]{iopart}

\usepackage{graphicx}
\usepackage{epsfig}
\usepackage{bm}

\begin{document}

\title[Proposed Experiment in Two-Qubit Linear Optical Photonic Gates]{Proposed Experiment in Two-Qubit Linear Optical Photonic Gates for Maximal Success Rates}

\author{A. Matthew Smith$^{1,2}$, D. B. Uskov$^{1,3}$, M. Fanto$^2$, L. H. Ying$^1$, and L. Kaplan$^1$}
\address{$^1$Department of Physics and Engineering Physics, Tulane University, New Orleans, Louisiana 70118,USA}%
\address{$^2$Air Force Research Lab, Information Directorate, 525 Brooks Road, Rome, New York 13440, USA}%
\address{$^3$Department of Mathematics and Natural Science, Brescia University, Owensboro, Kentucky 42301, USA}
\ead{amos.matthew.smith.ctr@rl.af.mil}

\begin{abstract}
Here we propose an experiment in Linear Optical Quantum Computing (LOQC) using the framework first developed by Knill, Laflamme, and Milburn.  This experiment will test the ideas of the authors' previous work on imperfect LOQC gates using number-resolving photon detectors.  We suggest a relatively simple physical apparatus capable of producing CZ gates with controllable fidelity less than 1 and success rates higher than the current theoretical maximum (S=2/27) for perfect fidelity.  These experimental setups are within the reach of many experimental groups and would provide an interesting experiment in photonic
 quantum computing.
\end{abstract}
%~\cite{moehring,internet}
%~\cite{klm}
%~\cite{uskov1, uskov2, uskov3}

%\pacs{03.67.Lx  42.50.Ex  42.50.Dv}
\submitto{\NJP}
\maketitle

\section{Introduction}
Optics remains one of the most promising methods for quantum information processing and computing, due to long photon decoherence times, the ease of photon manipulation, and the ability to transmit quantum state information over very large distances.  Optical states have also been proposed as possible buses between matter qubits~\cite{moehring,internet}.  It is therefore desirable to be able to manipulate states at the single-photon level via gate implementation.  Knill, Laflamme, and Milburn (KLM) significantly advanced the prospect of single-photon quantum computing in their seminal paper~\cite{klm}, in which they overcame the need for nonlinear interactions by using the inherent nonlinearity of photon measurements.  In this scheme, the computational system is combined with ancillary modes, and the gate operation is performed on the enlarged state space. The ancilla modes are measured with photon-number-resolving detectors, leaving the computational modes undisturbed and in the desired output state provided the measurement is successful. In our previous work~\cite{uskov1, uskov2, uskov3}, the authors have shown that a combination of analytical and numerical techniques may be used to design optimal linear optical transformations implementing two- and three-qubit entangling gates.  Here we show results for non-ideal gates and suggest an experiment to test them.

The probabilistic nature of quantum measurement implies a trade-off between the success rate of the operation (the probability of obtaining the desired measurement outcome for the ancillary modes) and the fidelity (the overlap between the actual and desired states of the computational system when the ancilla measurement is successful).  Previously, solutions were obtained that have the maximum possible ancilla measurement success probability given the constraint of perfect fidelity for a specified transformation~\cite{uskov1,uskov2}. In practical implementations, however, the goal of perfect fidelity may not always be desirable or even attainable.  We have therefore generalized our previous techniques to the case of imperfect fidelity, and investigated the above-mentioned trade-off between the fidelity and success of the linear optical transformations. It was found that for sufficiently small deviations from perfect fidelity, a single optimization parameter determines the relationship between fidelity and optimal success rate~\cite{uskov3}.

In section~\ref{sec-theory} of this work we briefly describe the theory behind our proposed experiment. In section~\ref{sec-exp} we describe a simple experimental apparatus capable of producing CZ gates with varying fidelity and success rates.  In section~\ref{sec-conc} we summarize the experimental requirements and provide concluding remarks.

\section{Theory}
\label{sec-theory}
%--------------------------------------------------------------------------------------------------------------------------------------------------------------------------------------------------
%A Linear Optical Quantum Computing (LOQC) measurement-assisted transformation, as suggested by KLM, is schematically illustrated in figure~\ref{figschematic}.
The input state to the experiment $|{\Psi}_{\rm in}^{\rm comp}\rangle\times| \Psi^{\rm ancilla}\rangle$ is a product of the computational state containing $M_c$ photons in $N_c$ modes, and an ancilla state containing $M_a$ photons in $N_a$ modes. The $N_c$ computational modes are those on which the actual gate is intended to act. 
Assuming dual-rail encoding, each qubit is represented by one and only one photon in two computational modes, so we have $M_c=N_c/2$. The ancilla state may in general be separable, entangled, or an ebit state carrying spatially distributed entanglement~\cite{wilde}, though here we propose using only a product state of single-photon and zero-photon ancillas, which are relatively simple to produce in an experimental setting.
%We also allow for an arbitrary number $N_v$ of vacuum input modes.

The linear optical device transforms the creation operator $a_i^{(\rm in)\dag}$ associated with each input mode $i$ to a  sum of creation operators $\sum_j U_{i,j} a_j^{(\rm out)\dag}$. Here $U$, which contains all physical properties of the device, is an $N \times N$ matrix, where $N=N_c+N_a$ is the total number of modes.  The total input state may be written as a superposition of  Fock states $|\Psi_{\rm in} \rangle = |n_{1}, n_2, \ldots, n_N \rangle$, where $n_i$ is the occupation number of the $i$-th input mode, and $\sum n_i=M_c+M_a=M$ is the total number of photons.  The input state is transformed as
\begin{equation}
\label{Eq:OutState}
|\Psi_{\rm out}\rangle = \hat{\Omega} |\Psi_{\rm in}\rangle = \prod_{i=1}^{N} \frac{1}{\sqrt{n_{i} !}} \left( \sum_{j=1}^N U_{i,j} a_{j}^{(\rm out)\dag} \right)^{n_i}  |0 \rangle \,.
\end{equation}
We note that
$\hat {\Omega}$ is a multivariate polynomial of degree $M$ in the elements $U_{i,j}$.

Once the transformation is complete, a measurement is applied to the $N_a$ ancillary modes.  In the case of a number-resolving photon-counting measurement, $\langle \Psi_{\rm measured}|=\langle k_{N_c+1}, k_{N_c+2}, \ldots, k_{N}| $, where $k_i$ is the number of photons measured in the $i$-th mode of the ancilla.  The resulting transformation of the computational state is a contraction quantum map  $|\Psi_{\rm out}^{\rm comp} \rangle = \hat{A} |\Psi_{\rm in}^{\rm comp} \rangle/\| \hat{A} |\Psi_{\rm in}^{\rm comp} \rangle\|$~\cite{kraus}, where $\hat{A}=\hat{A}(U)$ is defined by
\begin{equation}
\label{Eq:proj}
\hat{A} | \Psi_{\rm in}^{\rm comp} \rangle = \langle k_{N_c+1}, k_{N_c+2}, \ldots, k_{N}| {\hat{\Omega}} |\Psi_{\rm in}\rangle\,.
\end{equation}
The linear operator $\hat{A}$, which maps computational input states to computational output states, contains all the information of relevance to the transformation.

We define the fidelity as the probability that the desired target gate 
$\hat{A}^{\rm Tar}$ has been faithfully implemented on the computational modes given a successful measurement of the ancilla modes:
\begin{equation}
\label{Eq:F}
F(\hat A)=\frac{|{\rm Tr} \hat{A}^\dagger \hat{A}^{\rm Tar}|^2}{ 2^{M_c}{\rm Tr} \hat{A}^\dagger  \hat{A} } \,,
\end{equation}
since ${\rm Tr} \hat{A}^{{\rm Tar}\dagger} \hat{A}^{\rm Tar}= 2^{M_c}$ for a properly normalized target gate.
As we are interested in deviations from perfect fidelity, we define $\delta\equiv 1-F$ as our main parameter~\cite{uskov3}.

We define the success rate of the ancilla measurement to be given by an average over all computational input states, 
\begin{equation}
\label{Eq:S}
S(\hat A)={\rm Tr} (\hat{A}^\dagger \hat{A})/2^{M_c} \|U\|^{2M} \,
\end{equation}
for general complex $U$.  Note that $U$ need not be unitary, as any matrix can be made unitary via the unitary dilation technique by adding vacuum modes~\cite{knill,uskov2}. 
We also note that the Hilbert-Schmidt norm $\langle\hat{A} |\hat{A}\rangle = {\rm Tr} (\hat{A}^\dagger \hat{A})/2^{M_c}$, used in our definition of $S$, is bounded above by the square of the operator norm, $\| \hat A \|^2 \equiv (\| \hat{A}\|^{\rm Max})^2 = {\rm Max}(\langle \Psi_{\rm in}^{\rm comp} | {\hat A}^{\dag} \hat A| \Psi_{\rm in}^{\rm comp} \rangle)$, and below by $(\| \hat A \|^{\rm min})^2 = {\rm Min}  (\langle \Psi_{\rm in}^{\rm comp} | {\hat A}^{\dag} \hat A | \Psi_{\rm in}^{\rm comp} \rangle)$, where the maximum and minimum are taken over the set of properly normalized input states. 
%For normalization we use the Hilbert-Schmidt norm $\langle\hat{A} |\hat{A}\rangle $. It is easy to verify that $(\| \hat{A}\|^{\rm Min})^2 \leq 
%\langle\hat{A} |\hat{A}\rangle \leq (\| \hat{A}\|^{\rm Max})^2 $.
In the limit $F \rightarrow 1$, $ \| \hat{A}\|^{\rm Min}/\| \hat{A}\|^{\rm Max} \rightarrow 1$, and all definitions of the success rate coincide.
%and $S$ becomes a well-defined quantity equal to the Hilbert-Schmidt norm. we refer to $S$ as {\it the} success probability for general complex $U$. 

The optimization method we have developed maximizes the success probability $S$ for a given target transformation ${\hat A}^{\rm Tar}$, for given ancilla resources, and for a given fidelity level $F \le 1$.  This is mathematically equivalent to unconstrained maximization of the function $S+F/\epsilon$ in the space of all matrices $U$, where $1/\epsilon$ is a Lagrange multiplier. Here $\epsilon \to 0^+$ corresponds to maximizing the success probability while requiring perfect fidelity ($F=1$).  As $\epsilon$ is increased, the maximum of $S+F/\epsilon$ yields linear optics transformations that maximize the success $S$ as a function of the fidelity $F$. Given one transformation $U$ that (locally or globally) maximizes success $S$ for a given fidelity $F$, $\epsilon$ may be continuously varied to obtain a one-parameter family of optimal transformations, tracing out a curve in success-fidelity space.    Note that in general the members of these families need not be all unitary, however for some gates of interest, including the CZ gate, all members of the family are unitary.

\begin{figure}[htbp]
   \centerline{\includegraphics[width=0.6\textwidth]{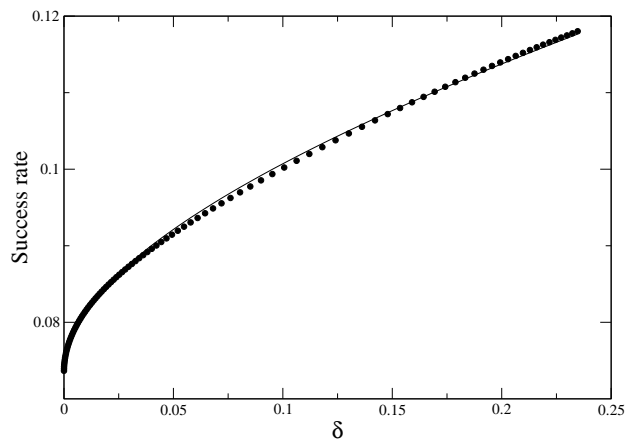} }
   \caption{{\bf Improved success rates for compromised $\delta$.}  Numerical results show the maximal success rate $S$ vs. $\delta$ for the CZ gate, and the curve is a best fit to the theoretical form (\ref{s-funct1}). Each data point rorrresponds to a distinct linear optical transformation, given by a unitary matrix.}
\label{cz-fig}
\end{figure}

Figure~\ref{cz-fig} shows optimal results for the CZ gate.  Here each point corresponds to a unique unitary mode transformation $U$.  As previously reported we find an interesting feature of these unitary matrices.  The optimal solution with fidelity $F=1$ was found by Knill to have a surprising form~\cite{knill}, which we have dubbed the ``Knill Form''~\cite{uskov2}, where one mode of each qubit is non-interacting, e.g., in the CZ case $U$ acts as the identity on modes $1$ and $3$ (or equivalently 1\&4, 2\&3, or 2\&4).  This form has been found to hold for the CZ gate and for the TS Toffoli Sign gate (CNOT and Toffoli respectively are equivalent to these up to local rotations).

We fit the data for the CZ gate to the known general analytic form \cite{uskov3},
\begin{equation}
\label{s-funct1}
S(F)=S_0+S_1(1-F)^{1/2}+S_2(1-F) +\dots \,,
\end{equation}
and truncate to
\begin{equation}
\label{s-numb-cnot}
S(\delta)=0.074+0.076 \,\delta^{1/2} \,.%+0.0457\delta \,.
\end{equation}
When $F=1$, $\delta=0$ and the the success $S$ reduces to the $2/27\approx 0.074$ result found by Knill~\cite{knill}, which has been confirmed numerically by the authors~\cite{uskov1}.  The ratio ${S_1/S_0}=1.03\pm.03$ contains the most interesting information about the system, as it is a measure of the relative rate of increase in success as the fidelity is compromised. We have used this ratio previously to compare the behavior of several gates, namely CNOT, CS(90), and the B gate~\cite{uskov3}.  All of these could be tested by the experimental apparatus proposed here, with some modifications.

\section{Experimental Design}
\label{sec-exp}
%--------------------------------------------------------------------------------------------------------------------------------------------------------------------------------------------------
We now propose an experiment that will test the results shown in figure~\ref{cz-fig}. Reck et al. have shown that any discrete $N \times N$ unitary transformation $U$ can be implemented as a multi-port device consisting only of variable transmittance beamsplitters and phase shifters~\cite{reck}.   Their method is a decomposition in which each unitary matrix element below the diagonal is transformed into zero by a $2 \times 2$ rotation matrix embedded in an $N \times N$ matrix which is otherwise equal to the identity. For example, the $2 \times 2$ rotation acting on modes $N$ and $N-1$, which eliminates the element $U_{N,N-1}$, takes the form
    \begin{equation}
\label{rot-mat}
T_{N,N-1}=\left(\begin{array}{cccc}
1&\dots&\dots&0\\ \vdots&\ddots&e^{i\phi}\sin(\omega)&e^{i\phi}\cos(\omega)\\0&\dots& \cos(\omega)&-\sin(\omega)\\ 
\end{array}\right) \,.
\end{equation}
The method is recursive and requires one iteration for each pair of modes. Finally, we obtain
\begin{equation}
\label{mats1}
U(N)T_{N,N-1}T_{N,N-2}\dots T_{2,1}D=I\,,
\end{equation}
where $D$ is a diagonal matrix of phases. The desired transformation $U$ is then decomposable as
\begin{equation}
\label{mats}
U(N)=D^{-1}T_{2,1}^{-1}T_{3,1}^{-1}\dots T_{N,N-1}\,.
\end{equation}
Physically, each $2 \times $2 transformation $T_{i,j}^{-1}$ is implemented as a variable transmittance beamsplitter with a phase plate on one input mode, while $D^{-1}$ corresponds physically to a phase shift on each output mode~\cite{reck}.

Thus a generic two-qubit operation, which needs at least $N=7$ modes ($N_c=4$ computational modes and $N_a=3$ ancillas) requires a minimum of $21$ beamsplitters and $28$ phase shifters.  A controlled unitary gate ($N=N_c+N_a=4+2=6$) requires at least $15$ beamsplitters and $21$ phase shifters.  If unitary dilation is required (as is often the case) the number of optical elements increases rapidly.  However our experiment does not require unitary dilation and furthermore as noted by Reck {\it et al.}, if an element of the unitary matrix is already zero, then no transformation is required.  The element is skipped.
      
\begin{figure}[htbp]
   \centerline{\includegraphics[width=0.6\textwidth]{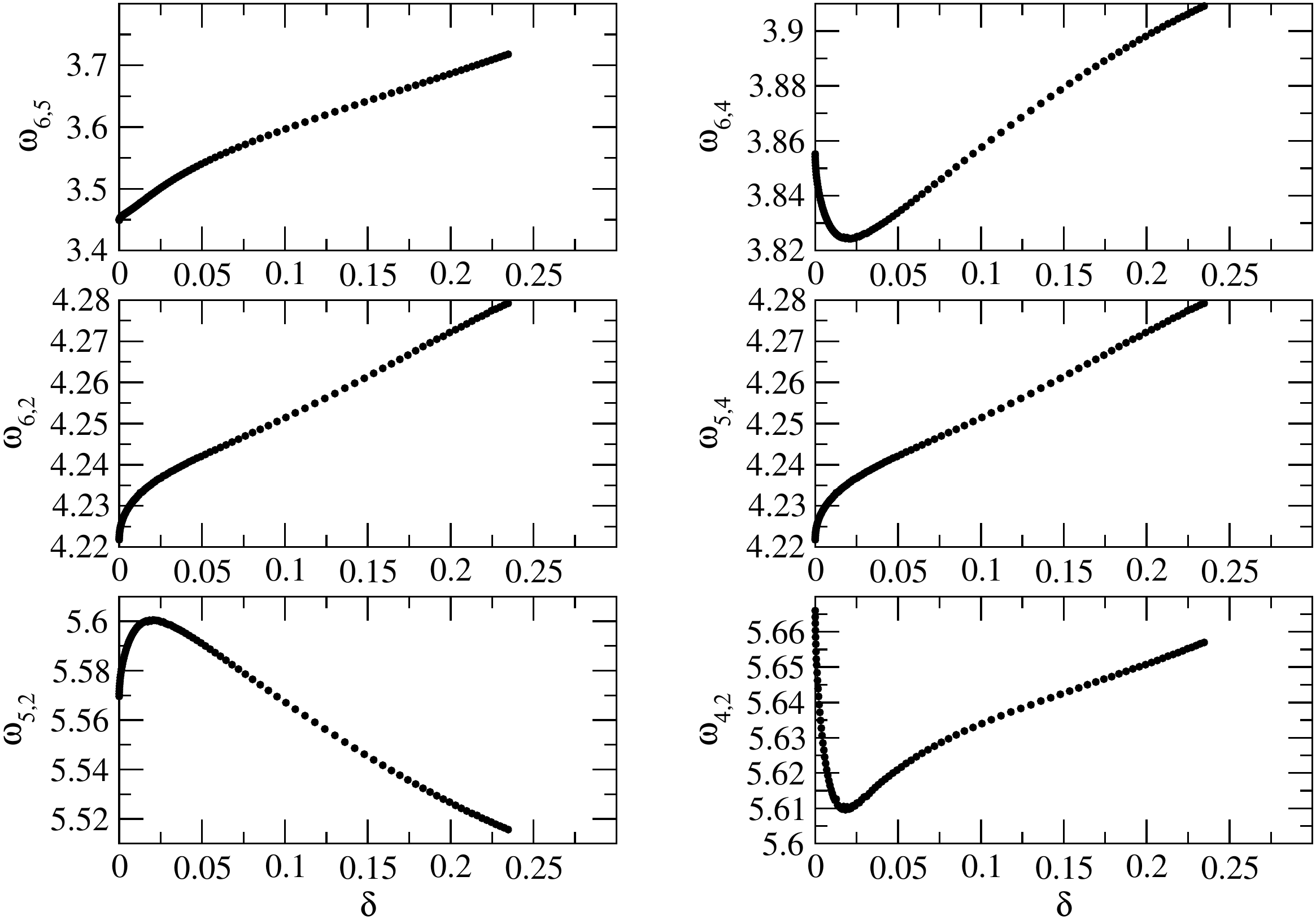}}
   \caption{{\bf Beamsplitter rotation angles.} Numerical results showing the six beamsplitter rotation angles $\omega_{i,j}$ vs. $\delta$ for the CZ gate.}
   %For each point delta the six beamsplitter transmission phases ($\omega$(6,5). . . $\omega$((4,2)) are shown.
\label{W}
\end{figure}

Here we return to the ``Knill Form,'' where in the case of CZ we find that nine of the elements below the diagonal are already zero. Therefore the unitary transform can be implemented with only six beamsplitters and 10 phase shifters.  We can perform this decomposition for each data point in figure~\ref{cz-fig}, and find the rotation angles $\omega_{i,j}$ and phases $\phi_{i,j}$ in each case. Surprisingly we find numerically that {\it all} of the phase shifts, $\phi_{i,j}$, are constant along the entire length of the curve in figure~\ref{cz-fig}. Therefore only the six beamsplitter rotation angles $\omega_{i,j}$ out of a total of 36 possible variables need to be modified to vary $\delta$, making the experiment much more physically realizable.  To be specific, the transformation only requires beamsplitters acting on the following mode pairs: $(i,j)=(6,5),\, (6,4), \, (6,2), \, (5,4), \, (5,2), \, (4,2)$.  Figure~\ref{W} shows that the six beamsplitter rotation angles change smoothly with $\delta$.  Implementing such rotations and constant phase shifters will recreate the unitary matrices from figure~\ref{cz-fig}.
%We find smooth curves for each of the 6 angles $\omega_{i,j}$.

This system lends itself to being implemented with $2 \times 2$ Mach-Zehnder interferometers (MZI) in place of standard beamsplitters.  The transmittance of the MZI is controlled dynamically by adjusting the phase difference, without having to alter the physical system.  These interferometers have already been put on optical chips by Thompson {\it et al.}~\cite{obrien2} among others.  Indeed, significantly larger electro-optical matrix switches have been proposed and built for broadband optical communication networks~\cite{obrien2,optnet}.   

\begin{figure}[htbp]
   \centerline{\includegraphics[width=0.6\textwidth]{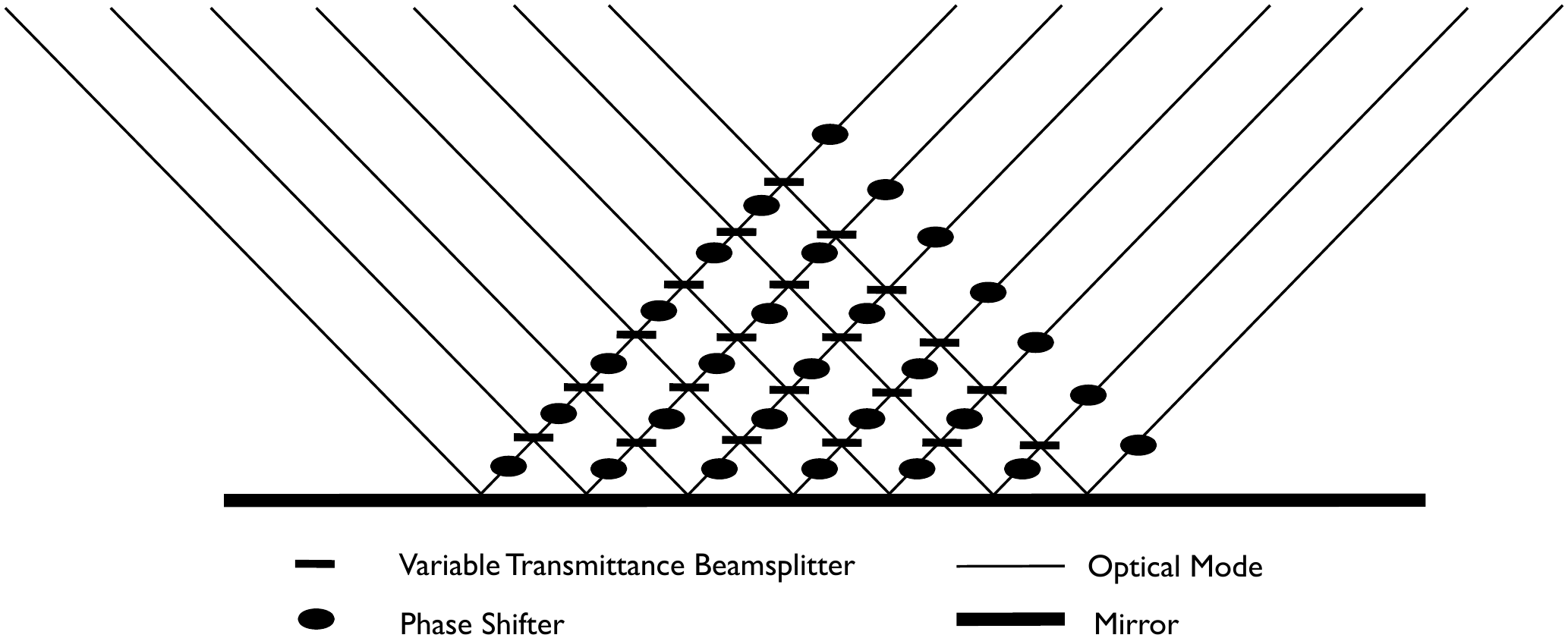}}
   \caption{{\bf General seven-mode optical device for two-qubit gates.} A schematic depiction of a multi-port optical device capable of performing our experiment as well as implementing any generic $SU(4)$ transformation in the measurement-assisted KLM scheme \cite{reck}.}
\label{mp}
\end{figure}

Figure~\ref{mp} shows a multi-port device that mixes seven input/output modes (thin lines) using $2 \times 2$ variable transmittance beamsplitters (rectangles), each of which has a phase shifter on one of its input modes (ellipses). An additional phase shifter is placed on each device output mode.  The thick line is a simple mirror.  %Such a device was described by \cite{reck}.
J. L. O'Brien recently proposed a similar $7 \times 7$ single-chip MZI-based device made from lithium niobate waveguides~\cite{obrien2}.
The intended purpose of this chip was to be able to perform any two-qubit unitary operation, i.e. any transformation in $SU(4)$.  However, such a device would also be capable of performing the experiment described above.

%  Not only reproducing the data shown in fig(\ref{cz-fig}) but confirming the key ratios shown here for CZ and published in \cite{uskov3} for CNOT.

Physical realization of work such as this can occur in many different experimental configurations.  A bulk optical component configuration allows the greatest flexibility in system reconfiguration.  The components required to construct this system are readily available at fairly low cost.  Unfortunately the system suffers from a large and badly scaling footprint size, thermal instabilities, and misalignment issues.  

The next logical step would be to implement this system in optical-fiber-based components, thus alleviating the need for critical alignment and reducing the overall footprint size.  Fiber allows for a reconfigurable system with properly connectorized components and drastically reduces the setup time for this experimental configuration.  The fiber-based components required for this experiment are also all readily available.  The phase shifters can be implemented in a manual stress-induced fiber phase shifter or a lithium niobate phase modulator.  The variable transmittance beamsplitters can be replaced with $2 \times 2$ evanescent couplers whose output port transmittance can be changed by manual adjustment of the gap between the fibers.  The tuning of these devices is slow; most are manually tuned and would require simultaneous tuning across the entire circuit.  Another possible component configuration consists of fiber-coupled bulk beamsplitters arranged to form a Mach-Zehnder interferometer with a lithium niobate phase shifter inserted into one of the arms. This would allow for high-speed and remote tuning of the transmittance.  Lastly, the beamsplitter could be implemented with a single device, a $2 \times 2$ lithium niobate MZI.   This device allows for the highest-speed reconfiguration and transmittance tuning of any of these possible suggested configurations.  The respective transmittance set for any particular MZI must be monitored due the thermal drift of the chip and the effect is amplified due to the modulators sinusoidal transfer function.  Many commercial MZI's address this problem with thermal stabilization of the chip and output power monitoring via an integrated photodetector.

However, fiber does pose a number of other challenges in a configuration such as this.  Polarization mode dispersion (PMD) occurring from random imperfections and asymmetries can lead to phase variations and these would need to be compensated for with the installation of a polarization controller.  PMD can be negated at the cost of accepting slightly higher fiber loss with the use of polarization-maintaining (strongly birefringent) fibers, in which the polarization is confined to one transmission plane. The other related issue to be aware of is polarization-dependent loss, again due to asymmetries, where one polarization experiences a higher loss rate.   Both of these effects can lead to transmittance differences through the interferometers in the experiment.   Thermal expansion can also induce phase variations in the fiber-based interferometers.  This can be actively compensated for with the addition of a Pound-Drever-Hall type feedback scheme, at the cost of increased system complexity~\cite{drever}.  Lastly, connector interfaces between components can lead to photon loss and state degradation.  

The ideal implementation would be a monolithic integrated circuit, which has a minimal footprint and provides greatest control of thermal stability.  There are a number of choices of material (silica, silicon, lithium niobate, etc.), but the final decision depends on the wavelength of interest, photon detection capabilities, and the intended function of the circuit. For example, silica will have lowest loss for passive waveguides in the telecom regime, but active devices in silica must be thermal and are therefore slow~\cite{matthews}. Lithium niobate will have higher loss waveguides than silica in the telecom regime, but it is far superior for active components such as modulators due to its nonlinear properties. 

Initial circuits including photon sources, waveguides, and detectors may have to be built on multiple chips, each with its own respective components that are best suited for being mated together and then eventually merged onto one monolithic chip.  Regardless of the chosen material, integrated waveguides allow movement towards scalable photon circuits~\cite{obrien2}.

Photon detection is the final piece of the puzzle.  Standard avalanche photodiodes (APDs) used with the detectors at 1550 nm have a higher noise than those at 800 nm, and neither are capable of photon-number resolution.  A second option is to use superconducting single-photon detectors, e.g. niobium nitride nanowire detectors~\cite{natarajan}, which can be configured to be capable of photon number resolution. The third option is to employ transition-edge detectors, which have demonstrated some of the highest number-resolving system detection efficiencies to date, at 98\%. 

\section{Conclusion}
\label{sec-conc}

We have shown the theoretical basis and interest for this experiment.  At this time it is the only apparent means of experimentally confirming the key ratio $S_1/S_0$, which quantifies the trade-off between fidelity and success, for the CZ or CNOT gate. The experimental setup may naturally be extended to explore the behavior of other quantum gates of interest. The components needed for the execution of the experiment are well within the means of many experimental groups.  The main stumbling block is the expense of purchasing number-resolving detectors.  However, any group already possessing these detectors should be able to implement this scheme with relative ease.

\section*{Acknowledgments}
This work was supported in part by the NSF under Grants PHY-0551164 and PHY-0545390, and by the National Research Council Research Associateship program at the AFRL Rome Research Site, Information Directorate.

%The authors declare that they have no competing financial interests.

\end{document}